%% file: TOP-15-001_temp.tex
\pdfoutput=1

\documentclass[11pt,twoside,a4paper,cmspaper,final,collab]{cms-tdr}
\def\svnVersion{407671}\def\cmsCernNoTag{CERN-EP/2017-012}\def\cmsCernDate{\today}\def\cmsMessage{Published in the European Physical Journal C as \href{http://dx.doi.org/10.1140/epjc/s10052-017-4912-8}{\doi{10.1140/epjc/s10052-017-4912-8.}}}

\begin{document}\cmsNoteHeader{TOP-15-001}

\hyphenation{had-ron-i-za-tion}
\hyphenation{cal-or-i-me-ter}
\hyphenation{de-vices}

\RCS$Revision: 403707 $
\RCS$HeadURL: svn+ssh://svn.cern.ch/reps/tdr2/papers/TOP-15-001/trunk/TOP-15-001.tex $
\RCS$Id: TOP-15-001.tex 403707 2017-05-09 22:52:35Z alverson $

 \newlength\cmsFigWidth
\ifthenelse{\boolean{cms@external}}{\setlength\cmsFigWidth{0.85\columnwidth}}{\setlength\cmsFigWidth{0.4\textwidth}}
\ifthenelse{\boolean{cms@external}}{\providecommand{\cmsLeft}{upper\xspace}}{\providecommand{\cmsLeft}{left\xspace}}
\ifthenelse{\boolean{cms@external}}{\providecommand{\cmsRight}{lower\xspace}}{\providecommand{\cmsRight}{right\xspace}}
\ifthenelse{\boolean{cms@external}}{\providecommand{\cmsLLeft}{Top\xspace}}{\providecommand{\cmsLLeft}{Left\xspace}}
\ifthenelse{\boolean{cms@external}}{\providecommand{\cmsRRight}{Lower\xspace}}{\providecommand{\cmsRRight}{Right\xspace}}

 \cmsNoteHeader{TOP-15-001}
 \title{Measurement of the top quark mass using single top quark events in proton-proton collisions at $\sqrt{s}= 8$\TeV}

 \date{\today}

 \abstract{
A measurement of the top quark mass is reported in events containing a single top quark produced via the electroweak $t$ channel.
The analysis is performed using data from proton-proton collisions collected with the CMS detector at the LHC at a centre-of-mass energy of 8\TeV, corresponding to an integrated luminosity of  19.7\fbinv.
Top quark candidates are reconstructed from their decay to a \PW\ boson and a b quark, with the \PW\ boson decaying leptonically to a muon and a neutrino. The final state signature and kinematic properties of single top quark events in the $t$ channel are used to enhance the purity of the sample, suppressing the contribution from top quark pair production. A fit to the invariant mass distribution of reconstructed top quark candidates yields a value of the top quark mass of $172.95 \pm 0.77\stat ^{+0.97}_{-0.93}\syst\GeV$. This result is in agreement with the current world average, and represents the first measurement of the top quark mass in event topologies not dominated by top quark pair production, therefore contributing to future averages with partially uncorrelated systematic uncertainties and a largely uncorrelated statistical uncertainty.
}

\hypersetup{
pdfauthor={CMS Collaboration},
pdftitle={Measurement of the top quark mass using single top quark events in proton-proton collisions at sqrt(s)= 8 TeV},
pdfsubject={CMS},
pdfkeywords={CMS, single-top, mass}}

\maketitle

 \section{Introduction}

{\tolerance=800
All previously published measurements of the top quark mass have been obtained using samples of top quark-antiquark pairs. A combination of measurements from the CDF and \DZERO\ experiments at the Tevatron and ATLAS and CMS experiments at the LHC yields a value of $173.34 \pm 0.27\stat \pm 0.71\syst\GeV$ for the top quark mass $m_{\PQt}$~\cite{TopMassCombo}.
Measuring $m_{\PQt}$ in single top quark production enriches the range of available measurements, exploiting a sample which is almost statistically independent from those used by previous ones, and with systematic uncertainties partially uncorrelated from those considered in \ttbar production.
Because of the different production mechanism, the mass extraction is affected differently by the modelling of both perturbative effects, such as initial- and final-state radiation, and nonperturbative effects, such as colour reconnection, in quantum chromodynamics (QCD).
Some discussion on these topics, though mainly restricted to the case of pair production, can be found in Refs.~\cite{Juste:2013dsa, Argyropoulos:2014zoa}.
In perspective, the lower level of gluon radiation and final state combinatorial arrangements with respect to \ttbar production will make this channel a good candidate for precision measurements of $m_{\PQt}$ when larger samples of events are available.
\par}

At the CERN LHC, top quarks are mainly produced as \ttbar pairs,  through gluon-gluon fusion or quark-antiquark annihilation, mediated by the strong interaction.
The standard model (SM) predicts single top quark production through electroweak processes, with a rate about one third that of the \ttbar production cross section. This has been confirmed by observations at the Tevatron ~\cite{Aaltonen:2013wca} and LHC ~\cite{Khachatryan:2014iya, Aad:2014fwa}.

In this paper, top quark candidates are reconstructed via their decay to a \PW\ boson and a \PQb quark, with the \PW\ boson decaying to a muon and a neutrino.
The event selection is tailored, before looking at data in the signal region, to enhance the single top quark content in the final sample and so have a result as independent as possible from those obtained using \ttbar events.

The paper is organised as follows.
Section 2 describes the CMS detector, followed by information about the data sample and simulation used in the analysis in Section 3.
The selection of events and the reconstruction of the top quark candidates is given in Section 4, and the description of the maximum-likelihood fit to derive the top quark mass is in Section 5.
Section 6 describes the systematic uncertainties affecting the measurement and Section 7 summarises the results.

\section{The CMS detector}
The central feature of the CMS apparatus is a superconducting solenoid of 6\unit{m} internal diameter, providing a magnetic field of 3.8\unit{T}. Within the solenoid volume are a silicon pixel and strip tracker, a lead tungstate crystal electromagnetic calorimeter, and a brass and scintillator hadron calorimeter, each composed of a barrel and two endcap sections. Forward calorimeters extend the pseudorapidity coverage provided by the barrel and endcap detectors. Muons are measured in gas-ionisation detectors embedded in the steel flux-return yoke outside the solenoid.

A more detailed description of the CMS detector, together with a definition of the coordinate system used and the relevant kinematic variables, can be found in Ref.~\cite{Chatrchyan:2008aa}.

\section{Data and simulated samples}

The measurement reported here is performed using the $\sqrt{s}=$8\TeV proton-proton collision data sample collected in 2012  with the CMS detector, corresponding to an integrated luminosity of 19.7\fbinv.

\begin{figure}
 \centering
     \includegraphics[width=0.32\textwidth]{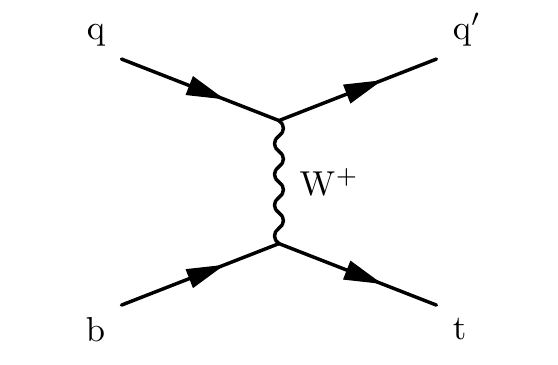}
     \includegraphics[width=0.32\textwidth]{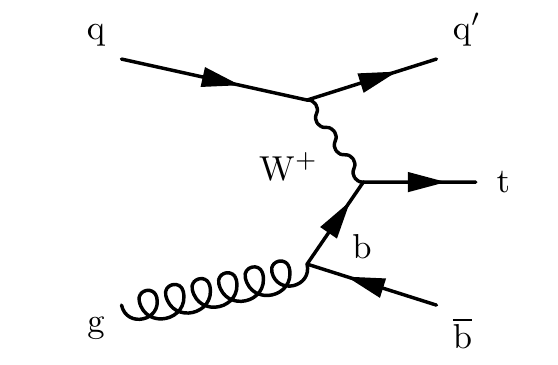}
   \hfill
   \caption{Feynman diagrams representing the dominant single top quark production mechanisms in the $t$ channel.}
    \label{fig:tchannel}
\end{figure}

{\tolerance=800
At the lowest order in perturbation theory, single top quark production proceeds through the $t$-channel, $s$-channel, and associated $\PQt\PW$ production modes.
The $t$ channel provides the largest contribution to the single top quark cross section.
The corresponding amplitude can be calculated using one of two different schemes~\cite{Campbell:2009gj, Maltoni:2012pa, Frederix:2012dh}: in the 5-flavour scheme, b quarks are considered as coming from the interacting proton, and the leading-order (LO) diagram is a $2 \to 2$ process (Fig.~\ref{fig:tchannel}, \cmsLeft); in the 4-flavour scheme, b quarks are not present in the initial state, and the LO diagram is a $2 \to 3$ process (Fig.~\ref{fig:tchannel}, \cmsRight).
The predicted $t$-channel single top quark cross section for pp collisions at a centre-of-mass energy of 8\TeV is $\sigma_{\PQt}= 54.9^{+2.3}_{-1.9}$\unit{pb} for the top quark and $\sigma_{\PAQt}=29.7^{+1.7}_{-1.5}$\unit{pb} for the top antiquark. These values are obtained by a next-to-leading-order (NLO) calculation in quantum QCD with \textsc{hathor} v.2.1~\cite{Aliev:2010zk, Kant:2014oha}, assuming a top quark mass of 172.5\GeV. The parton distribution functions (PDFs) and \alpS\ uncertainties are calculated using the PDF4LHC prescription~\cite{Alekhin, Botje:2011sn} with the MSTW2008 68\% confidence level (CL) NLO~\cite{Martin:2009iq, Martin:2009bu}, CT10 NLO~\cite{Lai:2010vv}, and NNPDF2.3~\cite{Ball:2012cx} PDF sets.
\par}

{\tolerance=800
At 8\TeV, the predicted \ttbar production cross section is $\sigma(\ttbar) = 252.9^{+6.4}_{-8.6}\,\text{ (scale)} \pm 11.7\,(\mathrm{PDF} + \alpS)\unit{pb}$ as calculated with the \textsc{Top++}2.0 program to next-to-next-to-leading order in perturbative QCD, including soft-gluon resummation to next-to-next-to-leading-log order (see Ref.~\cite{Czakon:2011xx} and references therein), and assuming a top quark mass of 172.5\GeV. In this calculation, the total scale uncertainty is obtained from the independent variation of the factorisation and renormalisation scales, $\mu_\mathrm{F}$ and $\mu_\mathrm{R}$, by a factor 2 and 1/2; the total PDF and $\alpS$ uncertainties are estimated following the PDF4LHC prescription~\cite{Botje:2011sn} with the MSTW2008 68\% CL NNLO~\cite{Martin:2009bu}, CT10 NNLO~\cite{Ball:2012cx}, and NNPDF2.3~\cite{Gao:2013xoa} FFN PDF sets.
\par}

Simulated events are used to optimise the event selection and to study the background processes and the expected performance of the analysis.
The  signal $t$-channel events are generated with the \POWHEG generator, version 1.0 \cite{Alioli:2010xd}, in the 5-flavour scheme, interfaced with \PYTHIA~\cite{Sjostrand:2006za}, version 6.426, for parton showering and hadronisation.
Single top quark $s$-channel and tW associated production are considered as backgrounds for this measurement and simulated with the same generator.
Top quark pair production, single vector boson production associated with jets (referred to as \PW/\PZ{}+jets in the following), and double vector boson (diboson) production are amongst the background processes taken into consideration and have been simulated with the \MADGRAPH generator, version 5.148 \cite{Alwall:2014hca}, interfaced with \PYTHIA for parton showering.
The \PYTHIA generator is used to simulate QCD multijet event samples enriched with isolated muons.
The value of the top quark mass used in all simulated samples is 172.5\GeV.
All samples are generated using the CTEQ6.6M \cite{Nadolsky:2008zw} PDF set and use the Z2* underlying event tune \cite{Z2Tune}.
The factorisation and renormalisation scales are both set to $m_{\mathrm{t}}$ for the single top quark samples, while a dynamic scale is used for the other samples, defined as the sum in quadrature of the transverse momentum (\pt) and the mass of the particles produced in the central process.
The passage of particles through the detector is simulated using the \GEANTfour toolkit \cite{Agostinelli:2002hh}.
The simulation includes additional overlapping pp collisions (pileup) with a multiplicity that is tuned to match the one observed in data.

\section{Event selection and reconstruction}

Signal events are characterised by a single isolated muon, momentum imbalance due to the presence of a neutrino, and one central b jet from the top quark decay. In addition, events often feature the presence of a light quark jet in the forward direction, from the hard-scattering process.

The online selection requires the presence of one isolated muon candidate with \pt greater than 24\GeV  and absolute value of the pseudorapidity ($\eta$) below 2.1.
Events are required to have at least one primary vertex reconstructed from at least four tracks, with a distance from the nominal beam-interaction point of less than 24 cm along the $z$ axis and less than 2 cm in the transverse plane.
In cases where more than one primary vertex is found, the one featuring the largest value of $\Sigma \pt^2$ is retained (``leading vertex''), where the sum runs over all the tracks assigned to that vertex.

All particles are reconstructed and identified with the CMS  particle-flow algorithm \cite{CMS:2009nxa, CMS:2010eua}.
Muon candidates are further required to have $\pt > 26\GeV$, thus ensuring they are selected in the region of maximal trigger efficiency.
Muon candidates are also required to be isolated.
This is ensured by requiring $I_{\text{rel}} < 0.12$, where $I_{\text{rel}}$ is defined as the sum of the transverse energies deposited by long-lived charged hadrons, photons, and neutral hadrons in a cone of size $\Delta R =\sqrt{\smash[b]{( \Delta \eta )^2 + ( \Delta\phi )^2}} = 0.4$ around the muon direction ($\phi$ being the azimuthal angle, in radians), divided by the muon \pt itself.
An offset correction is applied to remove the additional energy included in the jets that come from pileup \cite{CacciariPileup}.
Events are rejected if an additional muon (electron) candidate is present, passing the selection criteria $\pt> 10\, (20)$\GeV, $\abs{\eta}< 2.5$, and $I_{\text{rel}} < 0.2\ (0.15)$.

To define jets, the reconstructed particles are clustered using the anti-\kt algorithm~\cite{Cacciari:2008gp} with a distance parameter of 0.5.
Charged particles are excluded if they originate from a primary vertex that is not the leading vertex.
The energy deposition in the jet due to neutral pileup particles is inferred and subtracted by considering
charged pileup particles inside the jet cone.
Additional corrections to the jet energies are derived from the study of dijet events and photon+jets events \cite{Chatrchyan:2011ds}.
Jets are required to have $\abs{\eta} < 4.7$ and to have a corrected transverse energy greater than 40\GeV.
Jets associated with the hadronisation of b quarks ("b jets") are identified using a \PQb tagging algorithm based on the 3D impact parameter of the tracks in the jet to define a \PQb tagging discriminator~\cite{Chatrchyan:2012jua}.
The threshold for this variable is chosen such that the probability to misidentify jets coming from the hadronisation of light quarks (\PQu, \PQd, \PQs) or gluons is small (0.1\%), while ensuring  an efficiency of 46\% for selecting jets coming from b quarks, as determined from the simulation of events with top quark topologies.
Event weights are applied to adjust the b jet yields in the simulation to account for differences in the \PQb tagging efficiency between data and simulation.

The missing transverse momentum (\ptvecmiss) is calculated as  the negative vector sum  of the transverse momenta of all reconstructed particles. Corrections to the jet energies, as well as an offset correction accounting for pileup interactions, are propagated to \ptvecmiss.  The missing transverse momentum magnitude (\ptmiss) is required to exceed 50\GeV, to suppress the QCD multijet background.

To reject jets from pileup, non \PQb-tagged jets are rejected if the root-mean-square $\eta$-$\phi$ radius of the particles constituting the jet with respect to the jet axis is larger than 0.025.  To suppress background from QCD multijet events, the transverse mass of the \PW\ boson $m_\mathrm{T}(\PW)$ must be larger than 50\GeV, where $m_\mathrm{T}(\PW)$ is constructed from the missing transverse momentum and muon transverse momentum vectors as
\ifthenelse{\boolean{cms@external}}{
\begin{multline}
m_\mathrm{T}(\PW) =\\
 \sqrt{\left(\pt^{\mu} + \ptmiss \right)^2  - \left( p_x^{\mu} + p^{\text{miss}}_{\mathrm{T},x} \right)^2 - \left( p_y^{\mu} + p^{\text{miss}}_{\mathrm{T},y} \right)^2}.
\end{multline}
}{
\begin{equation}
m_\mathrm{T}(\PW) = \sqrt{\left(\pt^{\mu} + \ptmiss \right)^2  - \left( p_x^{\mu} + p^{\text{miss}}_{\mathrm{T},x} \right)^2 - \left( p_y^{\mu} + p^{\text{miss}}_{\mathrm{T},y} \right)^2}.
\end{equation}
}

The same event reconstruction and selection of top quark candidates adopted by the CMS single top quark $t$-channel cross section measurement at 8\TeV in Ref.~\cite{Khachatryan:2014iya} is used.
Due to the detector acceptance and jet selection requirements, most signal events are characterised by the presence of two reconstructed jets, one of which comes from the hadronisation of a \PQb quark.
Therefore, events with two reconstructed jets, exactly one of which is b tagged, constitute the ``signal region'' (referred to as `2J1T' in the following).
Other event topologies are used to study background properties: the sample with two reconstructed jets, neither of which is b tagged (`2J0T') is dominated by \PW+jets events; the sample with three reconstructed jets, where two jets are b tagged (`3J2T') is dominated by \ttbar events.
For all topologies considered, the jet with the highest value of the \PQb tagging discriminator is used for top quark reconstruction, while that with the lowest value is taken to be the light-quark jet associated with top quark production (Fig.~\ref{fig:tchannel}).

To enrich the sample in single top quark events, further requirements are applied to variables that exhibit good discriminating power with respect to \ttbar events, as described below.
The selection criteria have been chosen after studying their effect on the purity of the sample, while verifying that the statistical uncertainty achievable on the top quark mass would not be excessively degraded.

\begin{figure}
 \centering
   \includegraphics[width=0.49\textwidth]{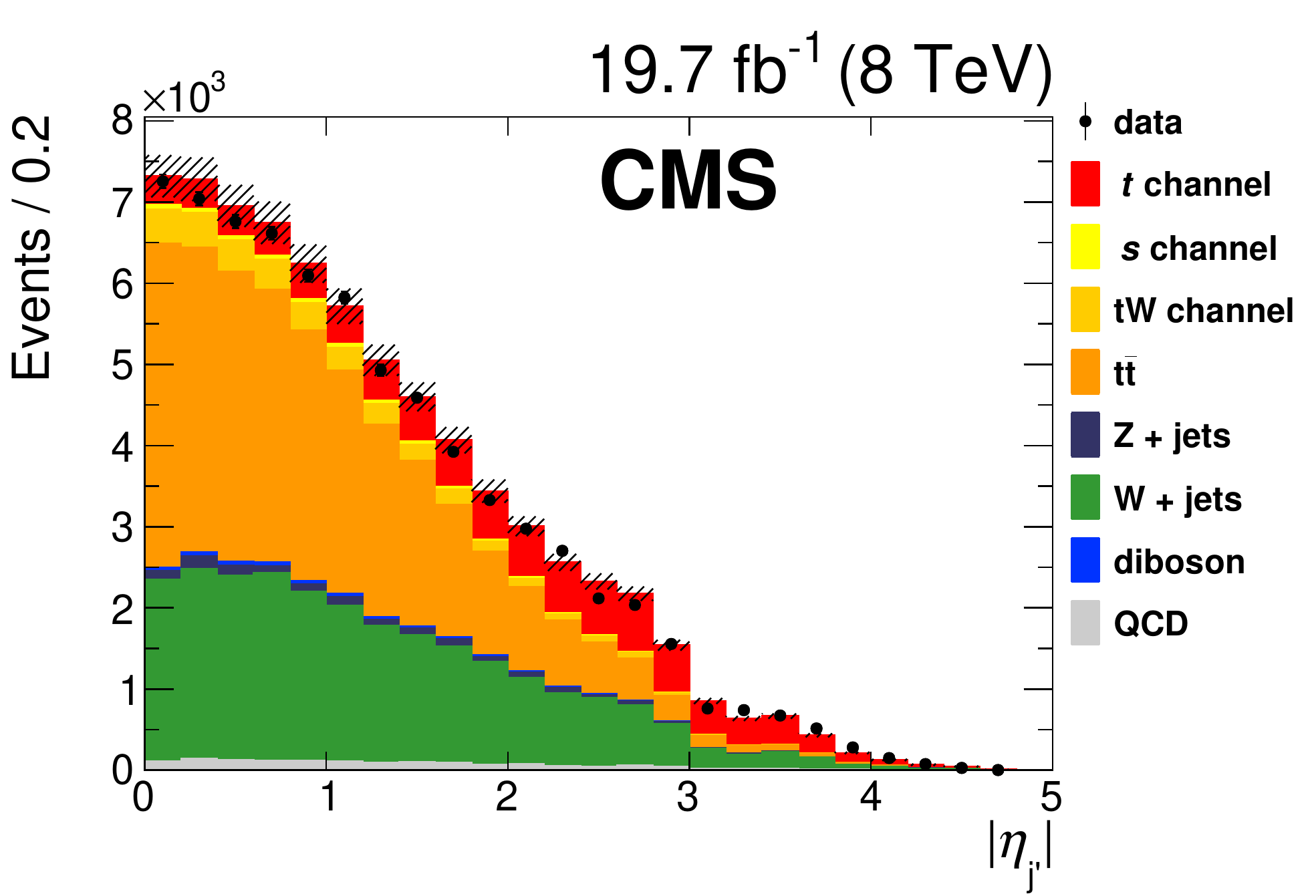}
   \includegraphics[width=0.49\textwidth]{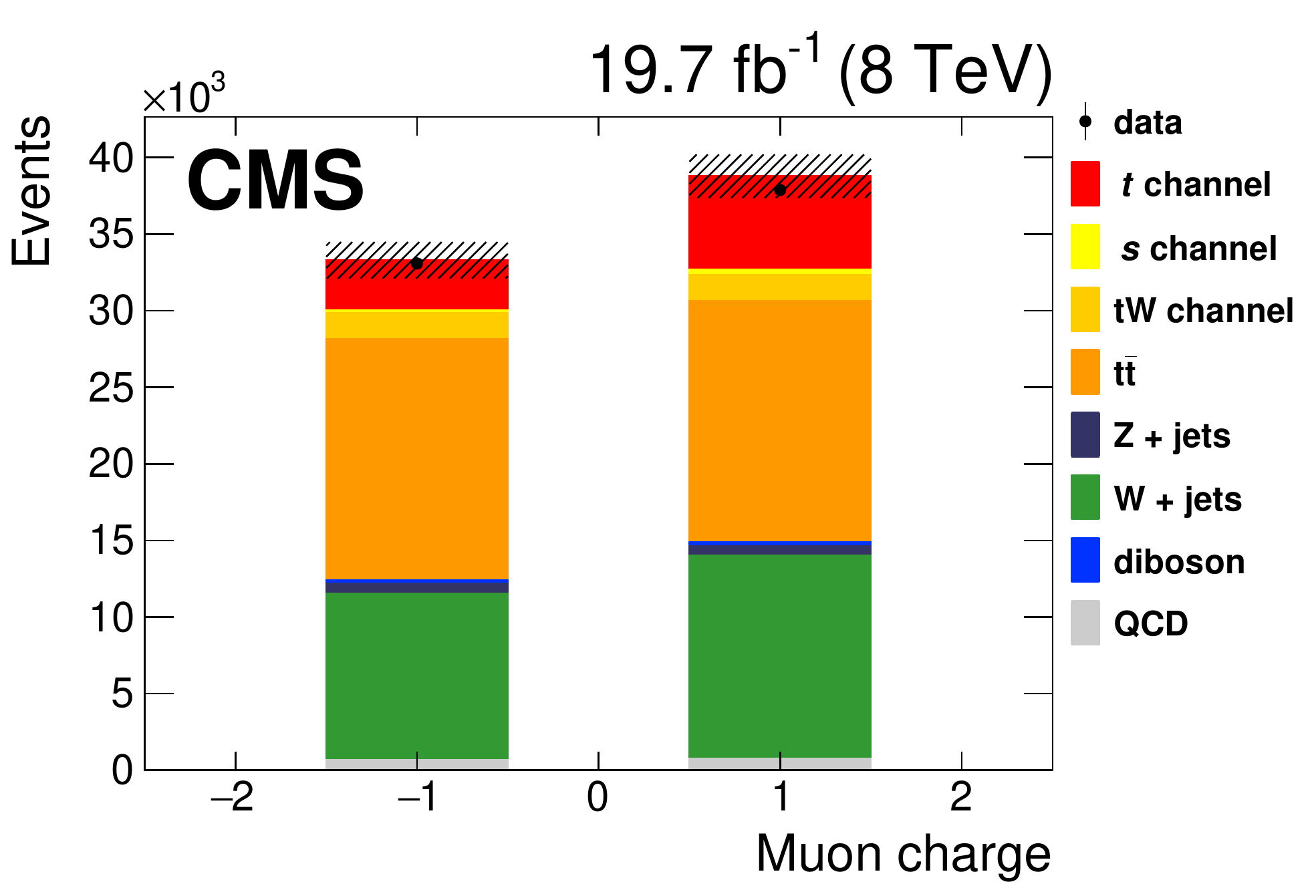}
    \caption{Distribution of the light-quark jet pseudorapidity (\cmsLeft) and of the muon charge (\cmsRight) for all top quark candidates in the muonic decay channel. Points with error bars represent data, stacked histograms show expected contributions from Monte Carlo simulation. The hatched area represents the uncertainty on the Monte Carlo predictions associated to the finite size of the samples and their normalization, and the integrated luminosity.}
    \label{fig:eta_chargeDistr}

\end{figure}

A feature of single top quark production in the $t$ channel is that the top quark is accompanied by a light-quark jet (the quark labelled $\PQq'$ in Fig.~\ref{fig:tchannel}), which is produced in a more forward direction than jets coming from \ttbar production or other background processes.
This is reflected in the distribution of the absolute value of the pseudorapidity of the light-quark jet $\abs{\eta_{\mathrm{j}'}}$, shown in Fig.~\ref{fig:eta_chargeDistr} (\cmsLeft) for all reconstructed top quark candidates. A requirement of $\abs{\eta_{\mathrm{j}'}} > 2.5$ is applied to the sample.
The stability of the selection has been checked by verifying that, if the events with $\abs{\eta_{\mathrm{j}'}} > 4$ were excluded, the final result would not be affected.

In $t$-channel single top quark production, top quarks are produced more frequently than top antiquarks due to the charge asymmetry of the proton-proton initial state \cite{Kidonakis:2012db}, as seen in the muon charge distribution (Fig.~\ref{fig:eta_chargeDistr}, \cmsRight).
To obtain as pure a sample as possible, only events with positively charged muons are retained.

\section{Determination of the top quark mass}

For each selected event, the top quark mass is reconstructed from the invariant mass $m_{\mu \nu \PQb}$ calculated from the muon, the neutrino, and the \PQb jet.
The 4-momenta of the muon and the jet are measured, while, for the neutrino, the 4-momentum is determined by using the missing transverse momentum in the event and a kinematical constraint on the \Pgm \PGn invariant mass, required to be consistent with the mass $m_{\PW}$ of the \PW\ boson~\cite{PDG2016}:
\ifthenelse{\boolean{cms@external}}{
\begin{multline}
m_{\PW}^2 = \left(E^{\mu} + \sqrt{(\ptmiss)^2+(p^{\nu}_{z})^2} \right)^2  \\
- \left( p_x^{\mu} + p^{\text{miss}}_{\mathrm{T},x} \right)^2 - \left( p_y^{\mu} + p^{\text{miss}}_{\mathrm{T},y} \right)^2-\left( p_{z}^{\mu} + p_{z}^{\nu} \right)^2 ,\label{eqn:neutrino}
\end{multline}
}{
\begin{equation}
m_{\PW}^2 = \left(E^{\mu} + \sqrt{(\ptmiss)^2+(p^{\nu}_{z})^2} \right)^2  - \left( p_x^{\mu} + p^{\text{miss}}_{\mathrm{T},x} \right)^2 - \left( p_y^{\mu} + p^{\text{miss}}_{\mathrm{T},y} \right)^2-\left( p_{z}^{\mu} + p_{z}^{\nu} \right)^2 ,\label{eqn:neutrino}
\end{equation}
}
where $E^{\mu}$ is the muon energy, $p_{x}^{\mu}$, $p_{y}^{\mu}$ and $p_{z}^{\mu}$ are the components of the muon momentum, $p_{z}^{\nu}$ is the longitudinal component of the neutrino momentum, and \ptmiss\ is used for the transverse components of the neutrino momentum. Equation \ref{eqn:neutrino} is quadratic in $p_{z}^{\nu}$: when two real solutions are found, the one with the smallest value of $|p_{z}^{\nu}|$ is taken; in the case of complex solutions, the imaginary component is eliminated by modifying $p^{\text{miss}}_{\mathrm{T},x}$ and $p^{\text{miss}}_{\mathrm{T},y}$ independently, so as to give  $m_\mathrm{T}(\PW) = m_{\PW}$ ~\cite{Bauer:2010ssa}.

\begin{figure*}
 \centering
   \includegraphics[width=0.49\textwidth]{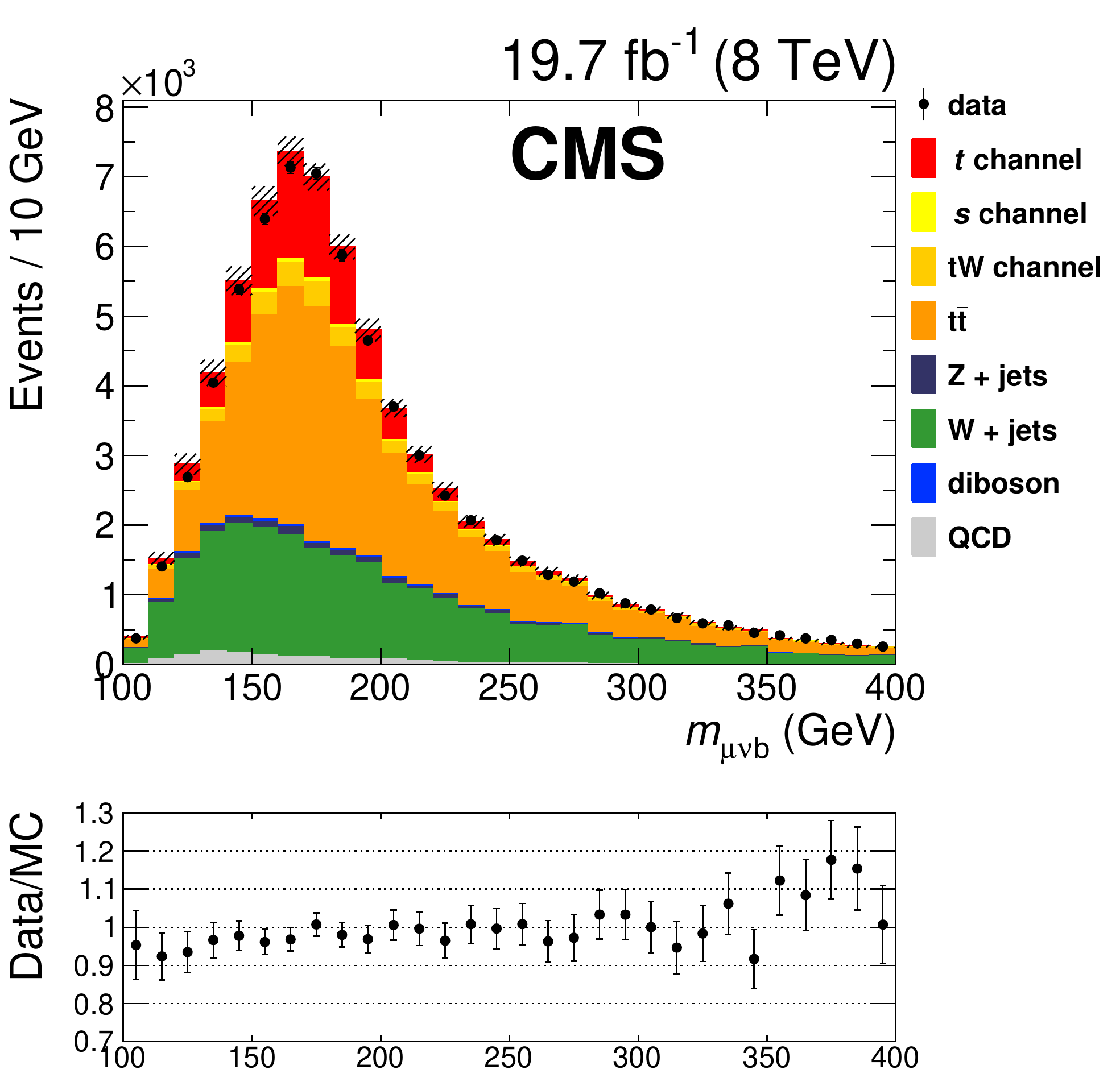}
   \includegraphics[width=0.49\textwidth]{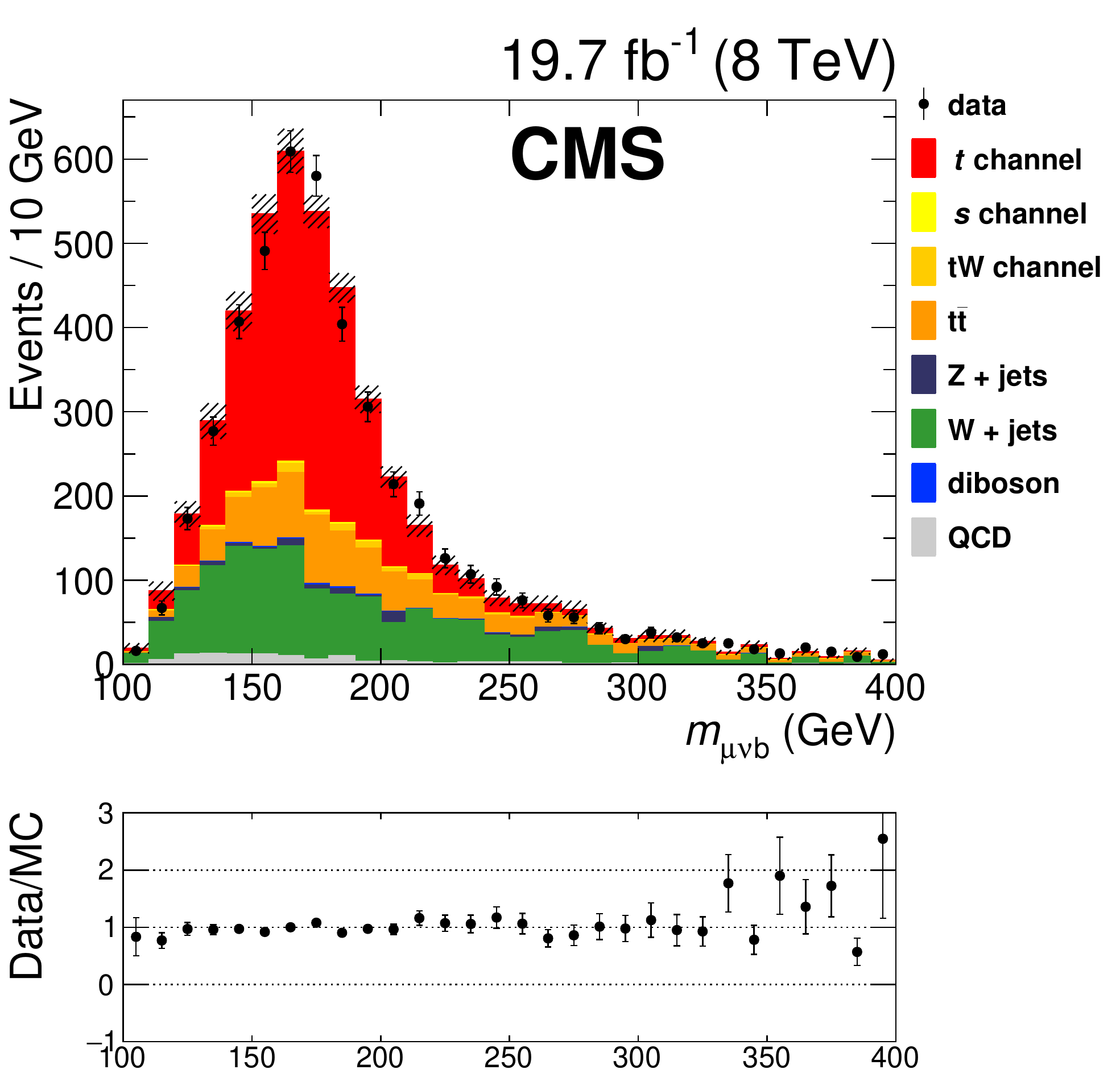}
    \caption{Reconstructed $\mu \nu \PQb$ invariant mass distribution for data (points with error bars) and Monte Carlo events (stacked histograms). Left: initial selection; right: final selection after the charge and light-quark jet pseudorapidity requirements. The ratio of the observed number of events in data to the number predicted by simulation is shown in the lower plots. The hatched area represents the uncertainty on the Monte Carlo predictions associated to the finite size of the samples and their normalization, and the integrated luminosity.}
    \label{fig:topMassDistr}
\end{figure*}

Figure~\ref{fig:topMassDistr} shows the $m_{\mu \nu \PQb}$ distributions  before and after the final event selection.
According to Monte Carlo simulation, after the final selection, 73\% of the reconstructed top quarks come from single top quark production, and of these about 97\% come from $t$-channel production.

The top quark mass is measured with an extended unbinned maximum-likelihood fit to the $m_{\mu \nu \PQb}$ distribution.
The numbers of events for the various contributions, except for the single top quark $t$-channel one, are fixed to the values extracted from simulation, taking into account the different theoretical cross sections~\cite{Khachatryan:2014iya}.
The description of the parameterisation of the signal and background components used in the fit is presented below.
The free parameters of the fit are the number of single top quark signal events and the parameters of the signal shape.

\subsection{Parameterisation of top quark components}
The shapes of the $m_{\mu \nu \PQb}$ distributions for samples where a top quark is present are studied using simulated events.

\begin{figure}
  \centering
    \includegraphics[width=0.49\textwidth]{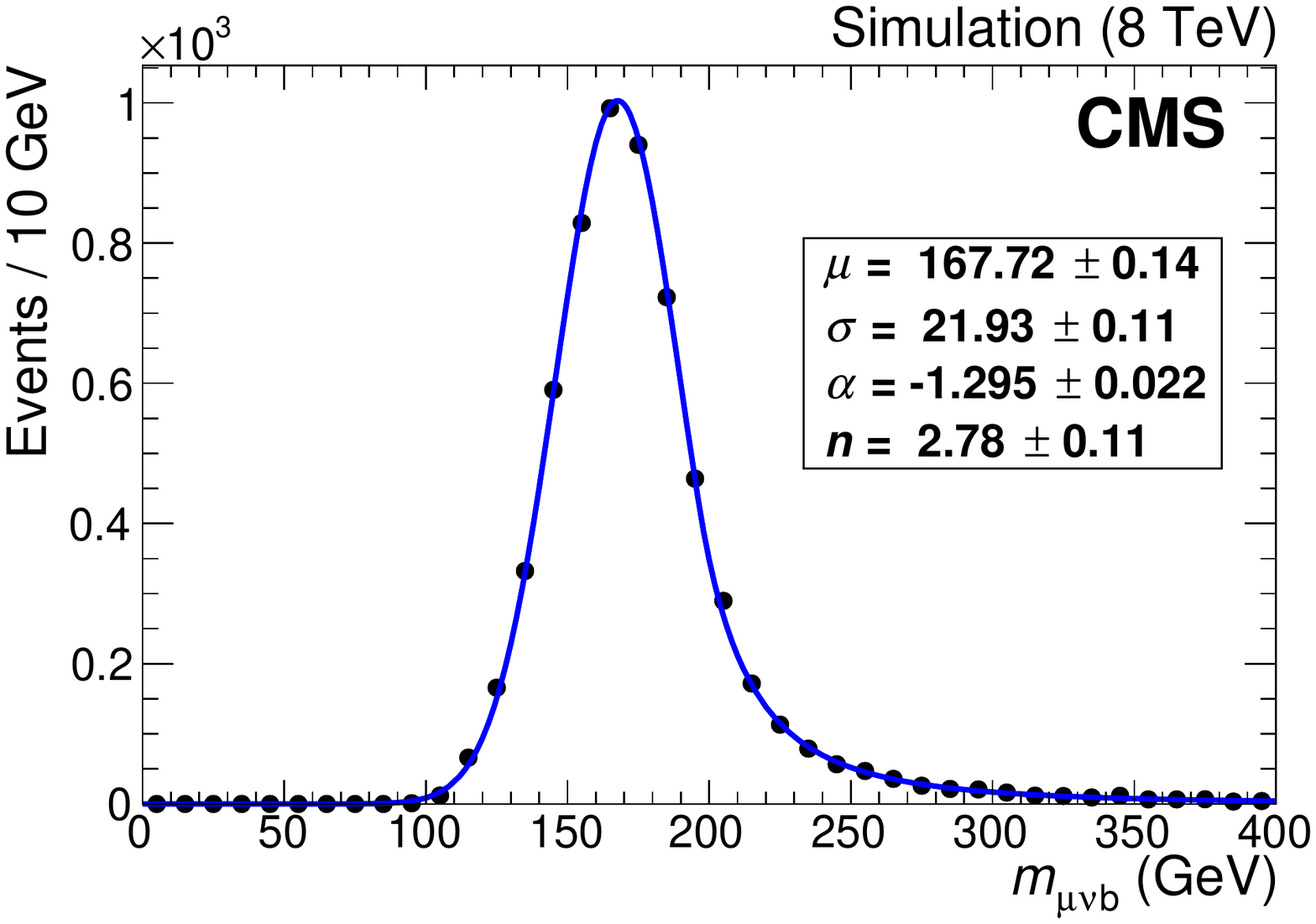}
    \includegraphics[width=0.49\textwidth]{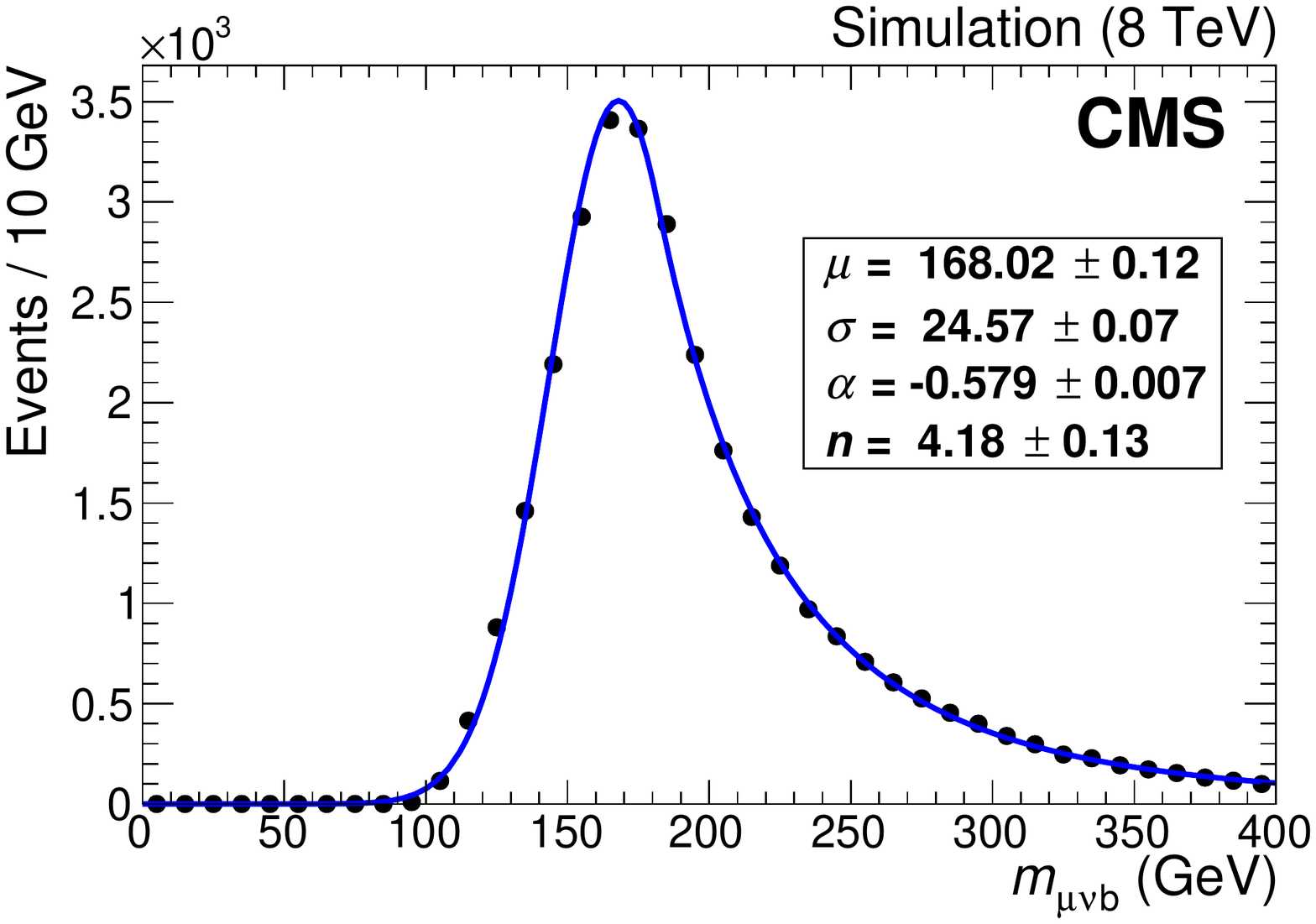}
    \caption{Reconstructed $\mu \nu \PQb$ invariant mass from Monte Carlo simulated events for single top quark $t$ channel (\cmsLeft) and \ttbar (\cmsRight). The continuous lines show the results of fits to Crystal Ball shapes.}
    \label{fig:signalShape}

\end{figure}

The \ttbar component exhibits a wider peak, with a larger high-mass tail, compared to the single top quark $t$-channel component.
The simulation shows that the number of muon and \PQb jet pairs correctly assigned to the parent top quark is around 55\% for \ttbar events, while this fraction exceeds 90\% for signal events.
Both contributions can be fitted by Crystal Ball functions~\cite{Oreglia}, with independent parameters $\mu$ and $\sigma$ representing the Gaussian core, and $\alpha$ and $n$ describing where the power-law tail begins and the exponent of the tail, respectively.
The distributions obtained from the simulated samples before the final selection are shown in Fig.~\ref{fig:signalShape}.
The difference between the values of the $\mu$ parameter of the Crystal Ball function obtained from the fits is $m_{\PQt}(\text{$t$ channel}) - m_{\PQt}(\ttbar) = 0.30 \pm 0.17\GeV$, where the uncertainty is the statistical uncertainty from the fit.

The remaining single top quark components ($s$-channel and $\PQt\PW$ production) account for only about 3.5\% of the final sample and their contribution is absorbed in the \ttbar component, since their distributions exhibit broader peaks with respect to the $t$ channel.

\begin{figure}
  \centering
    \includegraphics[width=0.49\textwidth]{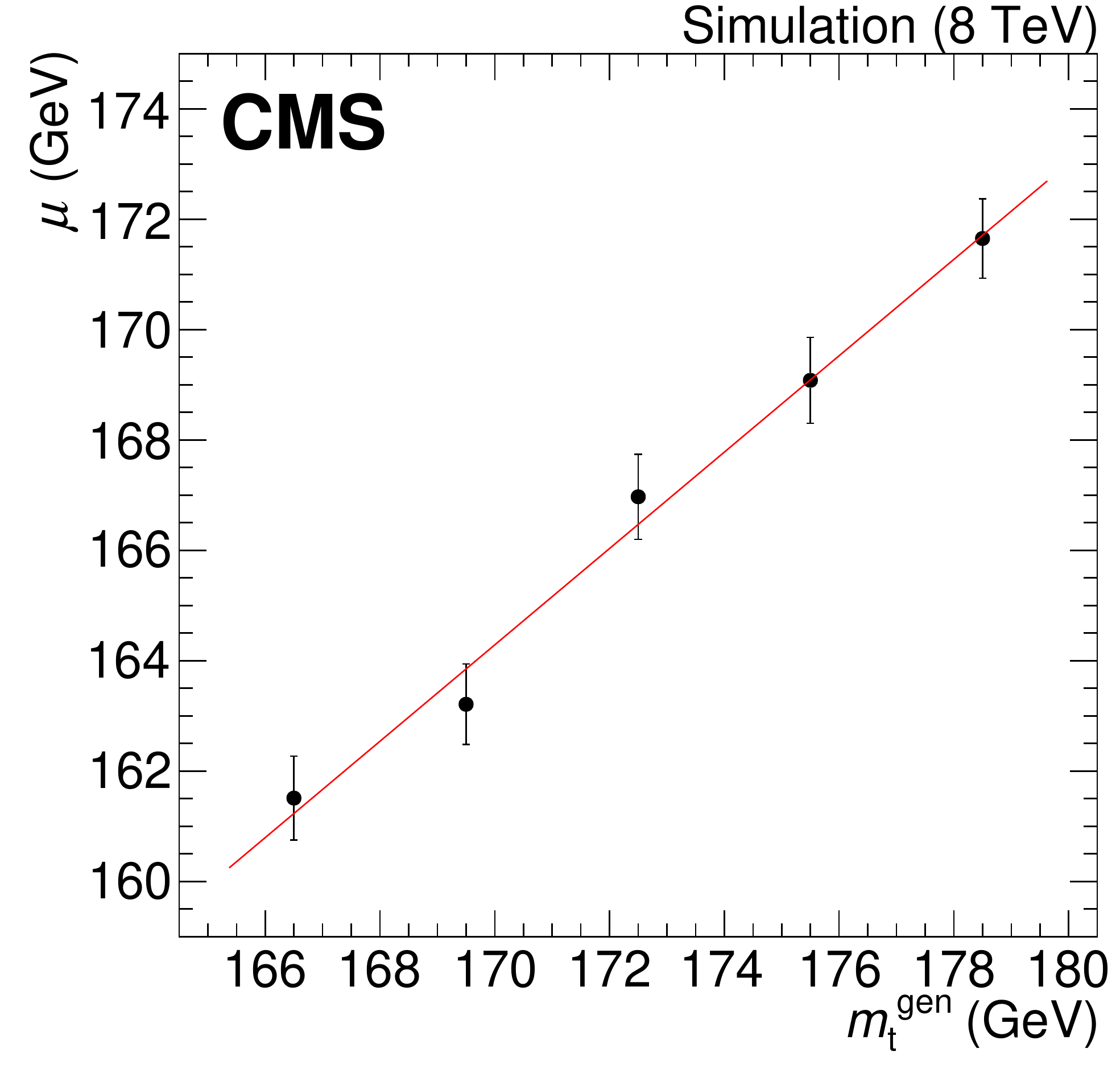}
    \includegraphics[width=0.49\textwidth]{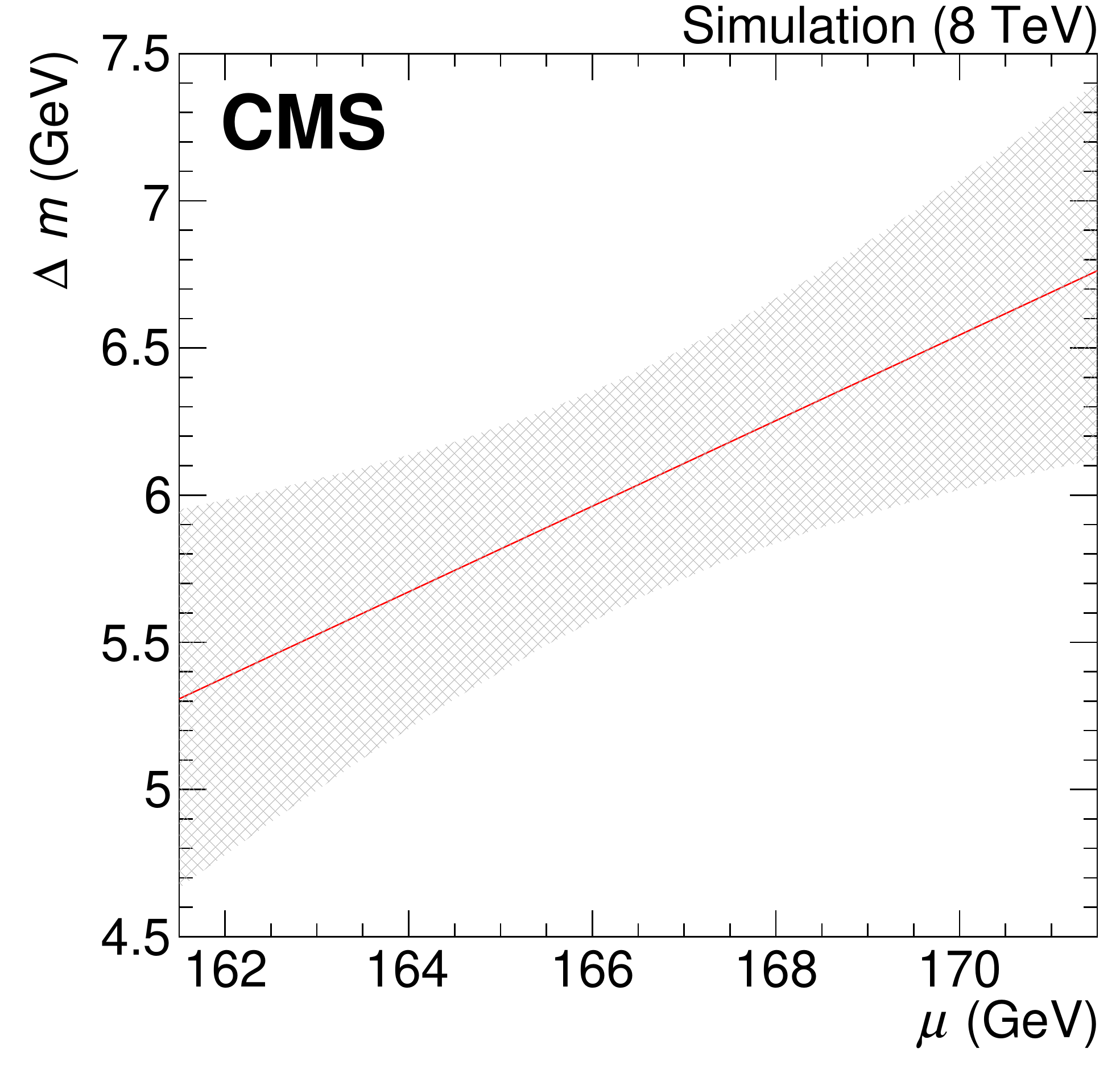}
    \caption{Mass calibration from fits to samples with different generated top quark mass. \cmsLLeft: fit results as a function of the generated top quark mass. The straight line shows the result of a linear fit to the chosen top quark mass values. \cmsRRight: mass shift, as a function of the fitted top quark mass (straight line). The shaded grey area represents the associated systematic uncertainty.}
    \label{fig:mTopCalib}

\end{figure}

The parameter $\mu$ of the Crystal Ball function describing the single top quark $t$-channel component is used to estimate the top quark mass.
The mass is obtained by shifting the value of $\mu$ resulting from the fit by an amount $\Delta m$ depending on $\mu$ itself.
In order to calibrate the magnitude of the shift, the fit has been repeated on a set of simulated samples including all signal and background processes, where the $t$-channel single top quark and \ttbar events were generated with different values of the top quark mass, all other events remaining unchanged.
Figure~\ref{fig:mTopCalib} shows the resulting values of $\mu$ as a function of the generated top quark mass (\cmsLeft) and the mass calibration curve from a fit to these values (\cmsRight).
The $\Delta m$ shift to be applied to the fitted value of $\mu$ is expressed as a linear function of $\mu$ itself.
The shaded grey area represents the uncertainty associated with $\Delta m$, obtained from the statistical uncertainties of the fits.

\subsection{Parameterisation of the non-top-quark background}

\begin{figure}
  \centering
    \includegraphics[width=0.49\textwidth]{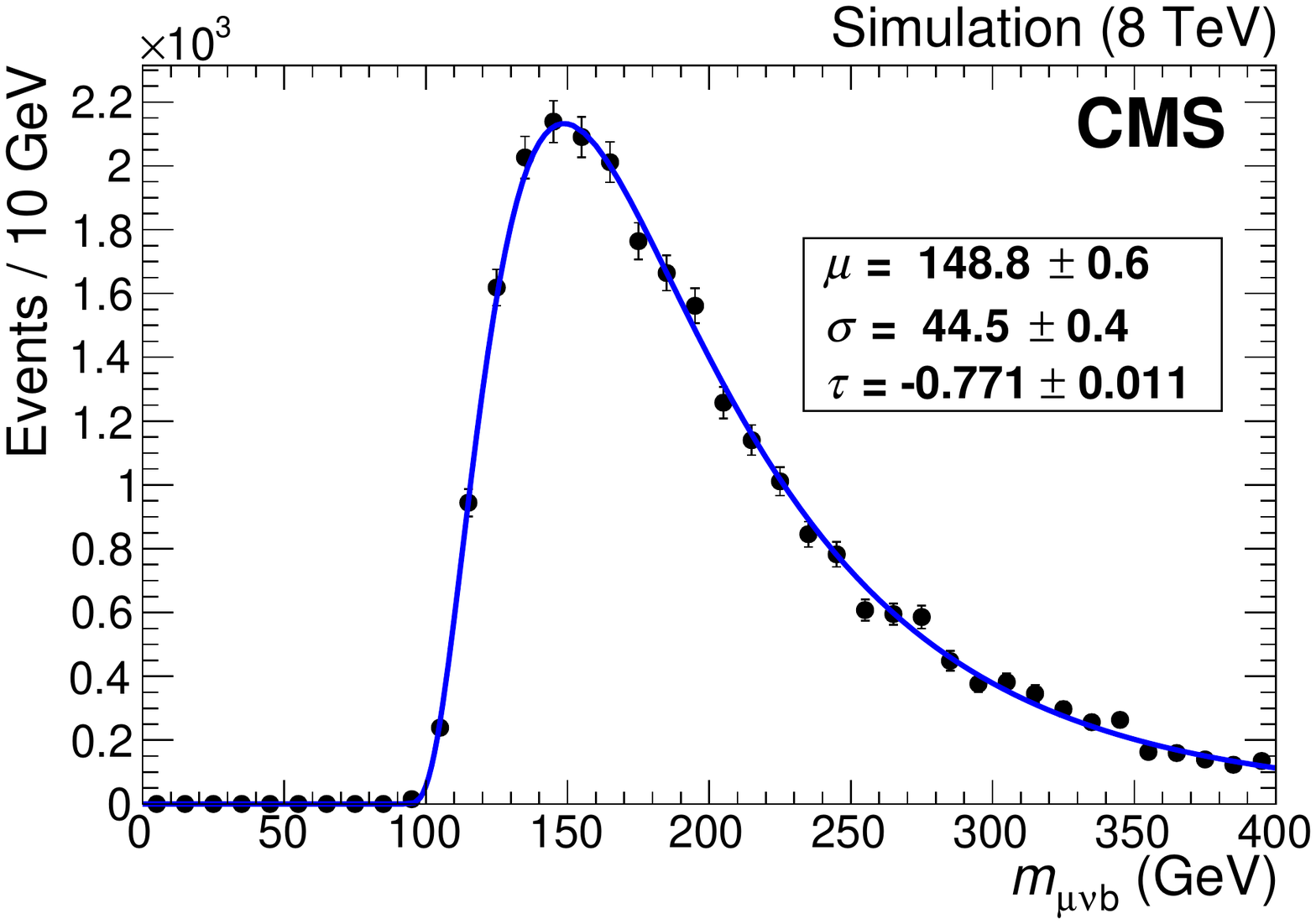}
    \includegraphics[width=0.49\textwidth]{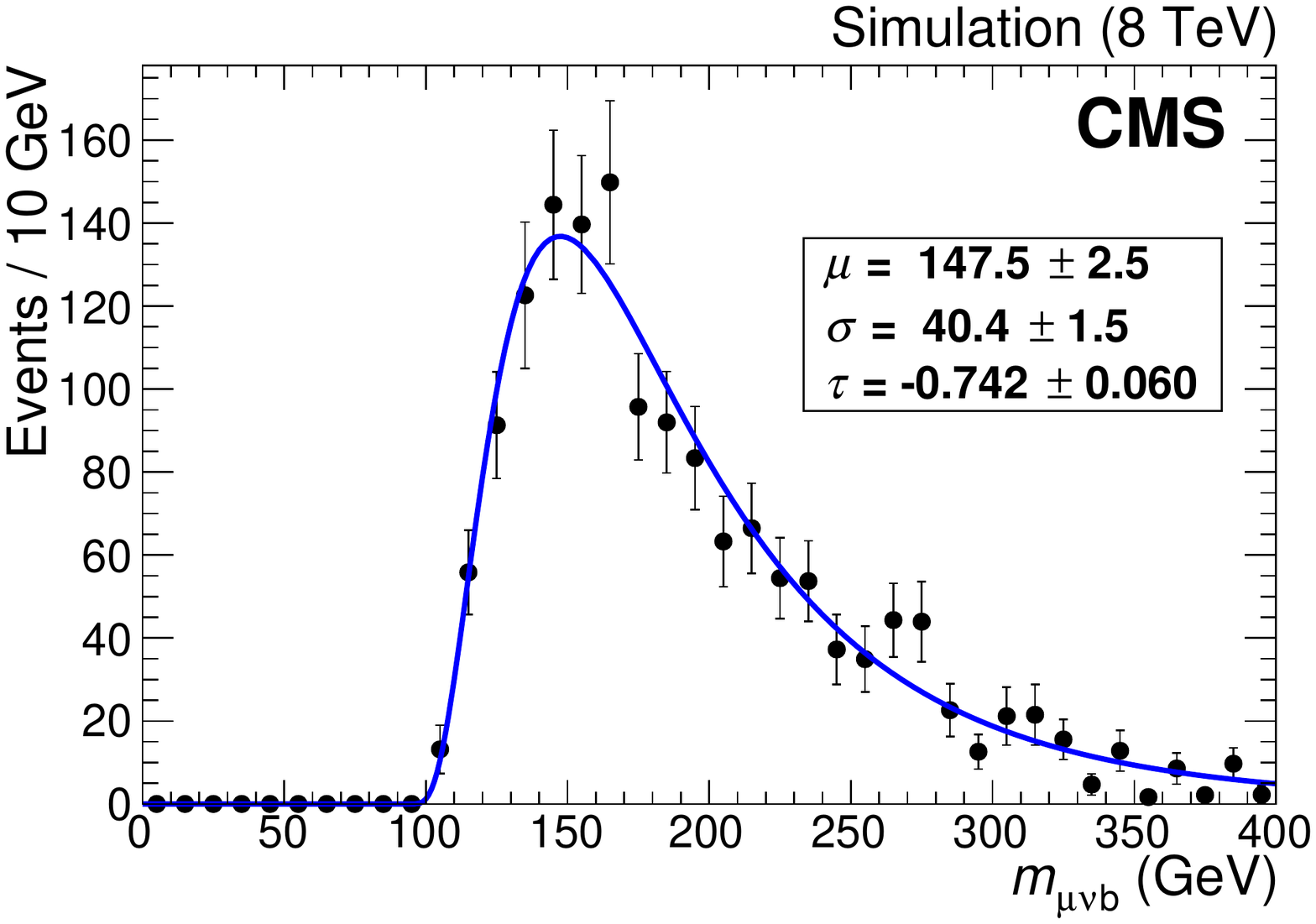}
    \caption{Reconstructed $\mu \nu \PQb$ invariant mass for non-top-quark background events, from Monte Carlo simulation. \cmsLLeft: before final selection; \cmsRight: after final selection. The continuous lines show the results of fits to Novosibirsk functions.}
    \label{fig:bkgShape}

\end{figure}

The \PW+jets events are expected to provide the largest contribution to the residual background.
The `2J0T' sample is mostly populated by such events and contains a large number of events, making it in principle a suitable control region to study the expected contribution of \PW+jets events to the background in the signal region.
However, the simulation shows that the reconstructed invariant mass distribution for \PW+jets events in the `2J0T' sample differs from that of the `2J1T' sample. Thus, simulation has been used for the characterisation of the \PW{}+jets component, as well as for all other non-top-quark background contributions.
The shape of the invariant mass distribution for the sum of all non-top-quark background sources is well reproduced by a Novosibirsk function~\cite{Ikeda:1999aq}, with parameters $\mu$ and $\sigma$ representing the Gaussian core, and $\tau$ describing the skewness of the distribution. The option to use the full simulated sample before the final selection, as is done for events containing top quarks, has not been chosen, as the parameters of the fitted function vary significantly with the requirement on  $\abs{\eta_{\mathrm{j}'}}$, as shown in Fig.~\ref{fig:bkgShape}.
Therefore, the sample obtained after applying the final selection is used to determine the shape parameters in the final fit.

\subsection{Determination of the top quark mass from the fit}

The invariant mass distribution of the selected top quark candidates is fitted with three components corresponding to signal, \ttbar and non-top-quark processes, using the probability density functions described above.
The mass is obtained from the resulting value of the mean of the Gaussian core of the Crystal Ball function fitting the single top quark contribution, applying the calibration procedure described above.
All parameters of the single top quark component are left free in the fit. The difference between the peak position of the $t$-channel and \ttbar components is kept fixed to the value measured in simulation, to reduce the statistical fluctuations due to the small number of residual \ttbar events. All remaining parameters (including normalisations) are fixed to the values extracted from simulation, after applying the final event selection.

\begin{figure}
  \centering
    \includegraphics[width=0.49\textwidth]{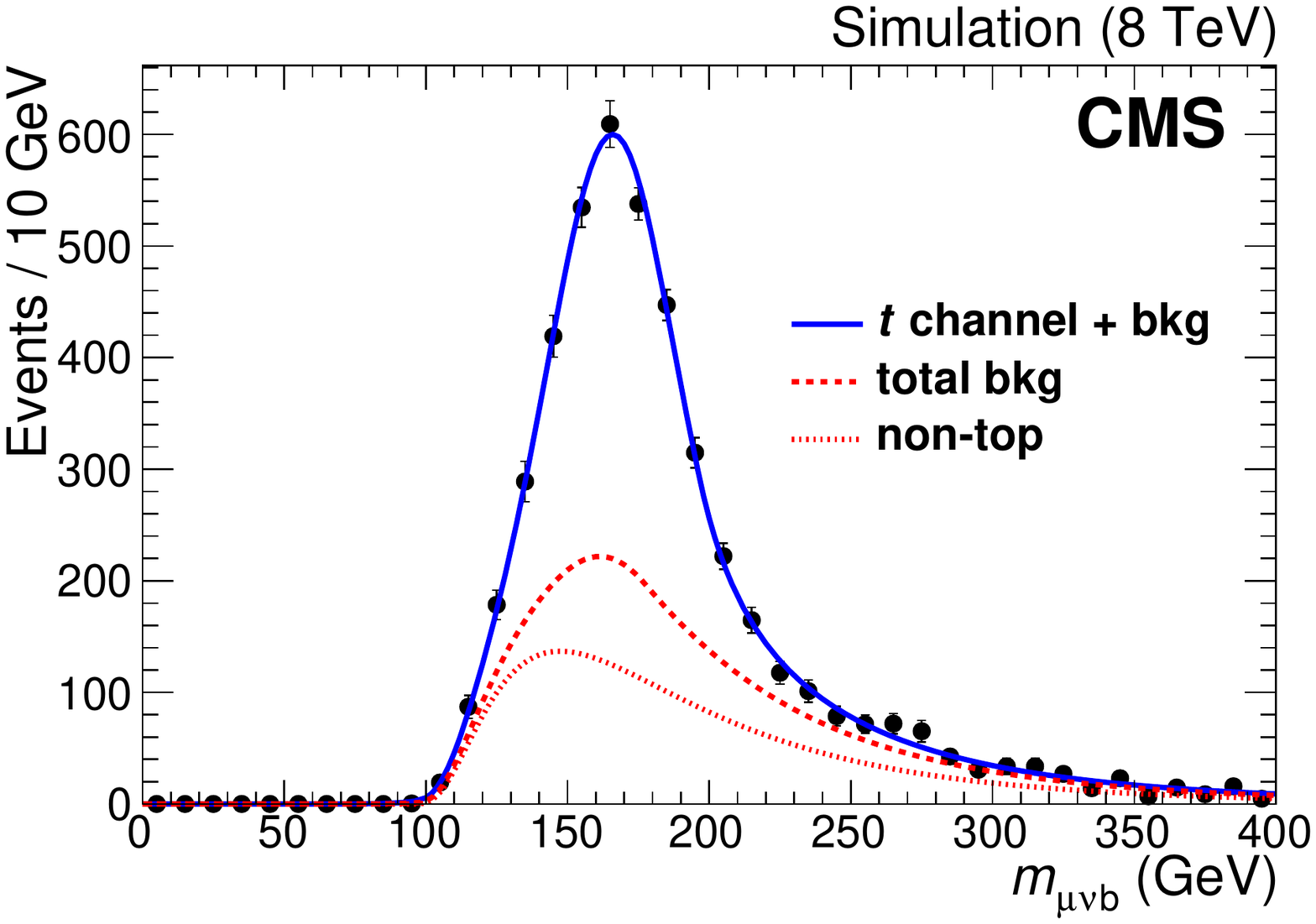}
    \includegraphics[width=0.49\textwidth]{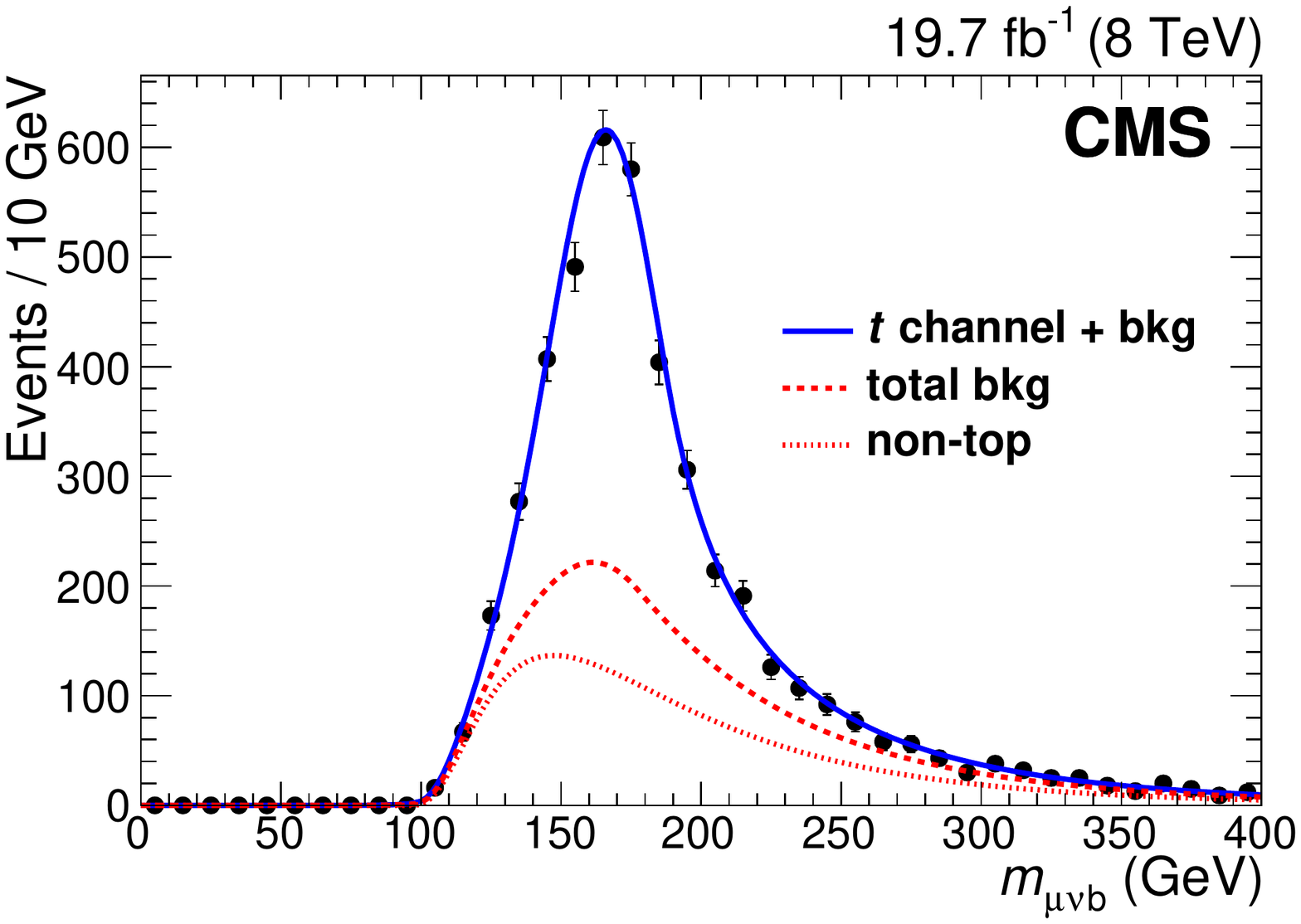}
    \caption{Result of the fit to the reconstructed $\mu \nu \PQb$ invariant mass. \cmsLLeft: Monte Carlo simulation; \cmsRight: data. In each plot, the solid line represents the result of the full fit; the dotted line shows the non-top-quark component, while the dashed line shows the sum of all background components.}
    \label{fig:finalFit}

\end{figure}

The results of the fits to the simulated sample and to the collision data sample are shown in Fig.~\ref{fig:finalFit}.
The number of $t$-channel events returned by the fit is $N_{\text{$t$-ch}}^{\text{fit}} = 2188 \pm 72$, in agreement with the number expected from simulation, $N_{\text{$t$-ch}}^{\mathrm{MC}} = 2216^{+94}_{-78}$.
A value of $m_{\PQt} = 172.95 \pm 0.77\stat\GeV$ is obtained after applying the mass calibration (Fig.~\ref{fig:mTopCalib}).
A systematic uncertainty of $0.39\GeV$ is associated to the mass calibration procedure.

\subsection{Cross-checks}

The consistency and stability of the fit are assessed using pseudo-experiments.
Ensembles of experiments are simulated using the signal and background templates, with their normalisations distributed according to Poisson statistics.
In each pseudo-experiment, the same fit described above is repeated and the top quark mass and the signal yield are derived.
The resulting distributions of the top quark mass and its root-mean-square show that the fit does not have any significant bias, with the difference between fitted and generated top quark masses, normalised to the fitted mass uncertainty (``pull''), distributed as expected.

Additionally, a test has been made where both the single top quark contribution and the \ttbar components are fitted with a single Crystal Ball function.
The results do not change appreciably within the present uncertainties with respect to the nominal fit.

The mass measurement for the single top quark contribution is derived after having removed the single top antiquark events.
As a check, the analysis has been repeated and the top quark mass has been measured using single top antiquark events.
The difference between the two measurements is $0.8 \pm 1.2\GeV$, with a difference of $-0.6 \pm 1.5\GeV$ expected from simulation.
Furthermore, the fit has been performed by simultaneously fitting single top quark and single top antiquark candidates: the fitted mass does not statistically differ with respect to the result obtained with the nominal fit.
These studies confirm that the selection of only the top quark candidates does not introduce any bias in the measured top quark mass.

\section{Systematic uncertainties}

Many of the uncertainties described below use modifications of the simulation to assess the impact on the final result.
These modifications affect the shapes and normalisations of the templates used by the fit.
Their contributions have been evaluated following the strategy adopted in Ref.~\cite{Khachatryan:2015hba}: the uncertainties are categorised consistently to allow effective combinations with other top quark mass measurements.

In the following, the sources of uncertainties identified as relevant for the measurement are described, as well as the procedure adopted to evaluate their impact. All the uncertainties are then combined in quadrature to derive the total systematic uncertainty.

\begin{description}

\item [{{Jet energy scale (JES):}}]

{\tolerance=800
JES factors are applied to the jet energy response in simulation to match that observed in data. The JES uncertainties are \pt- and $\eta$-dependent, and are taken into account by scaling the energies of all jets up and down according to their individual uncertainties, as determined in dedicated studies~\cite{Chatrchyan:2011ds}. The scaling is then propagated to the calculation of \ptmiss, and all other quantities dependent on the jet energies. The mass fit is repeated on the `scaled' simulated sample and the shift with respect to the nominal fit is taken as a measure of the uncertainty. The uncertainties in the JES are subdivided into independent sources and grouped into different categories following the prescription defined in Ref.~\cite{CMS:2015mvl}, aimed at simplifying the combination of measurements reported by the different LHC experiments. A total of 5 categories are identified referring to the effect of uncertainties related to the absolute scale determination using Drell--Yan events (``in-situ correlation group''), relative ($\eta$-dependent) calibration, and high- and low-\pt extrapolation (``inter-calibration group''), flavour-specific corrections (``flavour-correlation group''), pileup corrections using an offset dependence on the jet \pt (``pileup \pt uncertainty''), and remaining sources, uncorrelated between ATLAS and CMS (``uncorrelated group'').
\par}

\item [{{b quark hadronisation model:}}]

This is the term that accounts for the flavour-dependent uncertainties arising from the simulation of the parton fragmentation.

The total uncertainty can be decomposed into two separate contributions: the \PQb quark fragmentation uncertainty and the uncertainty from \PQb hadron decays.

The b quark fragmentation uncertainty has been derived in the same way as in the top quark mass measurement using semileptonic \ttbar events \cite{Khachatryan:2015hba}.
The Bowler--Lund fragmentation function for \PQb hadrons is retuned to agree with the $x_\PB$ data measured by the ALEPH~\cite{Heister:2001jg} and DELPHI~\cite{DELPHI:2011aa} Collaborations, where $x_\PB$ represents the fraction of the \PQb quark energy retained by a \PQb hadron.
A weight is attributed to each event, according to the $x_\PB$ value, and the difference with respect to the nominal setup is taken as the systematic uncertainty.

 The systematic uncertainty from the semileptonic branching ratio of \PQb hadrons is taken from Ref.~\cite{Khachatryan:2015hba}, in which the branching fractions were varied by $-0.45\%$ and $+0.77\%$ to give the possible range of values and the associated uncertainty.

\item [{{Jet energy resolution (JER):}}]

After correcting for the mismatch between the data and simulation for the energy resolution, the uncertainty is determined by varying the corrected JER within its $\eta$-dependent  $\pm $1 standard deviation uncertainties.
These changes are then propagated to the calculation of \ptmiss.

\item [{{Muon momentum scale:}}]

This contribution is determined by varying the reconstructed muon momenta by their uncertainties. These are determined  as a function of the muon $\eta$ and \pt with a ``tag-and-probe'' method based on Drell--Yan data, as described in Ref. \cite{LeptonScales}.

{\sloppy
\item [{{Unclustered missing transverse momentum:}}]

The uncertainty arising from the component of the missing transverse momentum that is not due to particles reconstructed as leptons and photons or clustered in jets (``unclustered \ptmiss'') is determined by varying it by $\pm$10\%.
\par}

\item [{{Pileup:}}]

This is the uncertainty coming from the modelling of the pileup in data.
It is taken as the sum of the effect of the uncertainty in the pileup rate (evaluated with pseudo-experiments in which the effective inelastic pp cross section has been varied by $\pm$5\%) and the pileup term extracted from the JES `uncorrelated' group (see above).

\item [{{b tagging efficiency:}}]

To calculate the uncertainties from the \PQb tagging efficiency and the misidentification rate, the \pt- and $\eta$-dependent \PQb tagging and misidentification scale factors are varied within their uncertainties for heavy- and light-flavour jets, as estimated from control samples \cite{Chatrchyan:2012jua}. The resulting changes are propagated to the event weights applied to the simulated events to obtain the uncertainties.

\item [{{Fit calibration:}}]

The mass is derived from the value of $\mu$ returned by the fit according to the mass calibration procedure described before: the same procedure provides an associated systematic uncertainty (Fig.~\ref{fig:mTopCalib}, \cmsRight).

\item [{{Background estimate:}}]

{\tolerance=800
This is the uncertainty resulting from the use of simulated backgrounds in the mass determination.
One contribution to the systematic uncertainty is determined by varying the background normalisations by $\pm$1 times their standard deviation uncertainties.
In addition, in the fit, the shape parameters of both the \ttbar and the W+jets components are fixed: these parameters are varied by $\pm$1 times their standard deviation uncertainties.
As there are theoretical uncertainties on the inputs to the simulation which may alter the background shapes used in the mass fit, an additional `radiation and matrix-element to parton-shower matching'  uncertainty is included, as described below.
\par}

\item [{{Generator model:}}]

Depending on whether the b quarks are considered part of the proton or not, the production of single top quarks can be studied in the 5- or 4-flavour schemes \cite{Frederix:2012dh}, respectively. The signal sample used in this work is produced with the 5-flavour scheme, where the b quarks are considered as constituents of the proton. To estimate the systematic uncertainty due to treating the b quarks like the lighter quarks, a comparison between a 5- and a 4-flavour-scheme ($2 \to 2$ and $2 \to 3$, respectively) samples has been performed: in the latter, the b quarks are generated in the hard scattering from gluon splitting. The samples used for the comparison are produced using the \COMPHEP generator \cite{Boos:2004kh}, version 4.5.1, with the same configuration as the nominal signal sample.

\item [{{Hadronisation model:}}]

This uncertainty is already covered by the JES uncertainty and \PQb quark hadronisation uncertainties considered above. As a cross-check, the nominal simulation is compared with results obtained using the \HERWIG generator~\cite{Herwig}, version 6.520, tune AUET2~\cite{ATLAS:2011gmi}, for parton showering and hadronisation. The resulting difference is in agreement with what is obtained for the JES uncertainty.

\item [{{Radiation and matrix-element to parton-shower matching:}}]

This is the category which covers the QCD factorisation and renormalisation scale (with the nominal values of $\mu_{\mathrm{R}}=\mu_{\mathrm{F}}=Q^2$) and initial- and final-state radiation uncertainties.

For the renormalisation and factorisation scale uncertainty determination, dedicated samples with $\mu_{\mathrm{R}}$ and $\mu_{\mathrm{F}}$ scales shifted up or down by a factor of 2 are used. The uncertainty is determined by comparing the central result with the shifted ones.

{\tolerance=800
For the matrix-element to parton-shower matching thresholds, \ttbar and \PW+jets samples in which the thresholds have been shifted up or down by a factor of 2 are used, with the systematic uncertainty evaluated in the same way as for the scale uncertainty. This is not relevant for the signal data set, which does not have a matrix-element to parton-shower matching. This procedure covers initial- and final-state radiation uncertainties.
\par}

All variations are applied independently of each other and the corresponding uncertainties are treated as uncorrelated.

\item [{{Underlying event:}}]

This term represents the uncertainty coming from the modelling of the underlying event (UE), the particles from the interaction that do not enter into the hard parton-parton interaction.
It is evaluated by comparing the results from two different tunes of \PYTHIA, the default Z2* tune and the Perugia tune \cite{Skands:2010ak}.
The differences in the value of the fitted mass are within the statistical uncertainty determined by the size of the simulated samples.
In fact, the two opposite variations result in mass shifts with the same sign. For this reason, the uncertainty from this source is estimated as the maximum statistical uncertainty of the changes.

\item [{{Colour reconnection:}}]

This uncertainty is evaluated by comparing two different UE tunes in which one has the nominal colour-reconnection effects and the other has these turned off.

\item [{{Parton distribution functions:}}]

The PDF4LHC \cite{Botje:2011sn} prescriptions are followed to calculate the uncertainty due to the choice of the PDFs.
The variation of the fitted top quark mass is estimated by using alternative sets of PDFs with respect to the nominal one, namely the MSTW2008CP \cite{Lai:2010vv}, CT10 \cite{Nadolsky:2008zw}, and NNPDF2.3 \cite{Martin:2009iq} sets.

\end{description}

The systematic uncertainties that affect the top quark mass measurement are summarised in Table~\ref{tab:syst}. Sources of systematic uncertainties that are totally or partially uncorrelated with  the top quark mass measurements using \ttbar\ events are the fit calibration, the background estimate, the generator model and theoretical parameters for the simulation of signal events, and the colour-reconnection effects.

\begin{table*}[htb]
\centering
\topcaption{Systematic uncertainties in the top quark mass.}
\begin{tabular}{llc}
 Source  & Subcategory & Uncertainty (\GeVns{})\\
\hline
\multirow{6}{*}{Jet energy scale} & In-situ correlation group & $+0.20, -0.21$\\
                                  & Inter-calibration group   & ${\pm}0.05$ \\
                                  & Flavour-correlation group & ${\pm}0.40$\\
                                  & Pileup \pt uncertainty &  $+0.18, -0.10$\\
                                  & Uncorrelated group        & $+0.48, -0.40$ \\
                                  \cline{2-3}
                                  & Total                     & $+0.68, -0.61$  \\[1ex]

 \PQb quark JES and hadronisation model  & & ${\pm}0.15$\\

 Jet energy resolution &  & ${\pm}0.05$\\

 Muon momentum scale & &  ${\pm}0.05$\\

 \ptmiss & &  ${\pm}0.15$\\

 Pileup & &  ${\pm}0.10$\\

 \PQb tagging efficiency  & & ${\pm}0.10$\\

 Fit calibration & & ${\pm}0.39$\\[1ex]

\multirow{5}{*}{Background estimate} & Shape          & ${\pm}0.10$ \\
                                     & Normalisation  & ${\pm}0.14$\\
                                     & $\mu_{\mathrm{R}}$ and $\mu_{\mathrm{F}}$ scales    & ${\pm}0.18$\\
                                     & Matching scales & ${\pm}0.30$\\
                                     \cline{2-3}
                                     & Total                     & ${\pm}0.39$\\[1ex]

Generator model  & & ${\pm}0.10$\\

 Signal $\mu_{\mathrm{R}}$ and $\mu_{\mathrm{F}}$ scales & &  ${\pm}0.23$  \\

 Underlying event & &  ${\pm}0.20$\\

 Colour reconnection & &  ${\pm}0.05$\\

 Parton distribution functions & &  ${\pm}0.05$\\
\hline
Total & & $+0.97, -0.93$
\end{tabular}
\label{tab:syst}
\end{table*}

\section{Summary}

The top quark mass is measured in a sample enriched in events with a single top quark for the first time.
Top quarks are reconstructed from decays to a \PW\ boson and a \PQb quark, with the \PW\ boson decaying to a muon and a neutrino.
In the final sample, events with a top quark from single production in the $t$-channel account for 73\% of the total number of events with a top quark.
The measurement is obtained from a fit to the distribution of the reconstructed mass of top quark candidates, where the $t$-channel single top quark component is modelled separately from the contribution of other top quark production channels.
The measured value is $m_{\PQt} = 172.95 \pm 0.77\stat ^{+0.97}_{-0.93}\syst\GeV$.
This is in agreement with the current combination of Tevatron and LHC results, $173.34 \pm 0.27\stat \pm 0.71\syst\GeV$, which is based on measurements using \ttbar events.
Because many of the systematic uncertainties affecting the measurement of $m_{\PQt}$ using single top quark $t$-channel events are totally or partially uncorrelated with the measurements using \ttbar events, and in addition the data sample analysed is largely statistically independent of the samples previously used, the result presented in this paper will be useful in the determination of world averages of the top quark mass.

\begin{acknowledgments}
We congratulate our colleagues in the CERN accelerator departments for the excellent performance of the LHC and thank the technical and administrative staffs at CERN and at other CMS institutes for their contributions to the success of the CMS effort. In addition, we gratefully acknowledge the computing centres and personnel of the Worldwide LHC Computing Grid for delivering so effectively the computing infrastructure essential to our analyses. Finally, we acknowledge the enduring support for the construction and operation of the LHC and the CMS detector provided by the following funding agencies: BMWFW and FWF (Austria); FNRS and FWO (Belgium); CNPq, CAPES, FAPERJ, and FAPESP (Brazil); MES (Bulgaria); CERN; CAS, MoST, and NSFC (China); COLCIENCIAS (Colombia); MSES and CSF (Croatia); RPF (Cyprus); SENESCYT (Ecuador); MoER, ERC IUT, and ERDF (Estonia); Academy of Finland, MEC, and HIP (Finland); CEA and CNRS/IN2P3 (France); BMBF, DFG, and HGF (Germany); GSRT (Greece); OTKA and NIH (Hungary); DAE and DST (India); IPM (Iran); SFI (Ireland); INFN (Italy); MSIP and NRF (Republic of Korea); LAS (Lithuania); MOE and UM (Malaysia); BUAP, CINVESTAV, CONACYT, LNS, SEP, and UASLP-FAI (Mexico); MBIE (New Zealand); PAEC (Pakistan); MSHE and NSC (Poland); FCT (Portugal); JINR (Dubna); MON, RosAtom, RAS, RFBR and RAEP (Russia); MESTD (Serbia); SEIDI, CPAN, PCTI and FEDER (Spain); Swiss Funding Agencies (Switzerland); MST (Taipei); ThEPCenter, IPST, STAR, and NSTDA (Thailand); TUBITAK and TAEK (Turkey); NASU and SFFR (Ukraine); STFC (United Kingdom); DOE and NSF (USA).

\hyphenation{Rachada-pisek} Individuals have received support from the Marie-Curie programme and the European Research Council and EPLANET (European Union); the Leventis Foundation; the A. P. Sloan Foundation; the Alexander von Humboldt Foundation; the Belgian Federal Science Policy Office; the Fonds pour la Formation \`a la Recherche dans l'Industrie et dans l'Agriculture (FRIA-Belgium); the Agentschap voor Innovatie door Wetenschap en Technologie (IWT-Belgium); the Ministry of Education, Youth and Sports (MEYS) of the Czech Republic; the Council of Science and Industrial Research, India; the HOMING PLUS programme of the Foundation for Polish Science, cofinanced from European Union, Regional Development Fund, the Mobility Plus programme of the Ministry of Science and Higher Education, the National Science Center (Poland), contracts Harmonia 2014/14/M/ST2/00428, Opus 2014/13/B/ST2/02543, 2014/15/B/ST2/03998, and 2015/19/B/ST2/02861, Sonata-bis 2012/07/E/ST2/01406; the National Priorities Research Program by Qatar National Research Fund; the Programa Clar\'in-COFUND del Principado de Asturias; the Thalis and Aristeia programmes cofinanced by EU-ESF and the Greek NSRF; the Rachadapisek Sompot Fund for Postdoctoral Fellowship, Chulalongkorn University and the Chulalongkorn Academic into Its 2nd Century Project Advancement Project (Thailand); and the Welch Foundation, contract C-1845.

This work is dedicated to the memory of Maurizio Lo Vetere (1965-2015), our fellow colleague at the University of Genoa and INFN, Italy. Maurizio's main contributions to this paper include the set up of the event selection to suppress \ttbar events and the signal extraction method.
\end{acknowledgments}

\bibliography{auto_generated}

\cleardoublepage \appendix\section{The CMS Collaboration \label{app:collab}}\begin{sloppypar}\hyphenpenalty=5000\widowpenalty=500\clubpenalty=5000\input{TOP-15-001-authorlist.tex}\end{sloppypar}
\end{document}

%% file: TOP-15-001-authorlist.tex
\textbf{Yerevan Physics Institute,  Yerevan,  Armenia}\\*[0pt]
A.M.~Sirunyan, A.~Tumasyan
\vskip\cmsinstskip
\textbf{Institut f\"{u}r Hochenergiephysik,  Wien,  Austria}\\*[0pt]
W.~Adam, E.~Asilar, T.~Bergauer, J.~Brandstetter, E.~Brondolin, M.~Dragicevic, J.~Er\"{o}, M.~Flechl, M.~Friedl, R.~Fr\"{u}hwirth\cmsAuthorMark{1}, V.M.~Ghete, C.~Hartl, N.~H\"{o}rmann, J.~Hrubec, M.~Jeitler\cmsAuthorMark{1}, A.~K\"{o}nig, I.~Kr\"{a}tschmer, D.~Liko, T.~Matsushita, I.~Mikulec, D.~Rabady, N.~Rad, B.~Rahbaran, H.~Rohringer, J.~Schieck\cmsAuthorMark{1}, J.~Strauss, W.~Waltenberger, C.-E.~Wulz\cmsAuthorMark{1}
% ...
\vskip\cmsinstskip
% ...
1:~~Also at Vienna University of Technology, Vienna, Austria\\
% ...
